\documentclass[app,twocolumn,showpacs,preprintnumbers,amsmath,amssymb]{revtex4}

\usepackage{amsmath,amssymb,multirow,epsfig,bm,color}

\newcommand{\beq}{\begin{equation}}
\newcommand{\eeq}{\end{equation}}
\newcommand{\bea}{\begin{eqnarray}}
\newcommand{\eea}{\end{eqnarray}}

\newcommand{\benn}{\begin{displaymath}}
\newcommand{\eenn}{\end{displaymath}}
\newcommand{\angstr}{\rm\AA}

\begin{document}

\title{Magnetic properties of two-dimensional M$_2$N$_3$ (M-metal, N=S,Se,Te) compounds.}

\author{K. Zberecki$^1$}
%\author{Author1$^1$, Author2$^1$, Author3$^1$}
\affiliation{$^1$Faculty of Physics, Warsaw University of Technology, ul. Koszykowa 75, 00-662 Warsaw, Poland}
%\affiliation{$^2$Faculty of Physics, Adam Mickiewicz University, ul. Umultowska 85, 61-614 Pozna\'n, Poland\\ and Institute of Molecular Physics, Polish Academy of Sciences, Smoluchowskiego 17, 60-179 Pozna\'n, Poland}
\date{\today}
\email{zberecki@if.pw.edu.pl} 

\begin{abstract}

Using \textit{ab-initio} methods we study structural, electronic and magnetic properties of two dimensional compounds with stoichiometry M$_2$N$_3$ (M-metal, N=S,Se,Te). Our study shows that structures with Cr, Ti, and Mn are stable, with significant binding energy. Also, such structures are semiconductors with narrow band gaps. 
We also show that Cr$_2$Se$_3$, Cr$_2$Te$_3$ and Mn$_2$Te$_3$ have considerable magnetic moments. The negative values of magnetic anisotropy energy suggest, that these materials can maintain ferromagnetic ordering in non-zero temperatures with estimated Curie temperatures in the range of 30-55 K. 
\end{abstract}

\pacs{...}

\maketitle

\section{Introduction} 

Two-dimensional (2D) ferromagnetic (FM) materials, are at present one of the most promising building blocks of nanoscale spintronic devices \cite{materi2D}. Very recently, FM Cr$_{2}$Ge$_{2}$Te$_{6}$ bilayer \cite{magn1}, as well as CrI$_{3}$ \cite{magn2} and Fe$_{3}$GeTe$_{2}$ \cite{magn3} monolayers have been successfully synthesized. However, the Curie temperature (T$_C$) of these intrinsic 2D FM materials lies far below room temperature because of the weak ferromagnetic super-exchange interaction, preventing them from most applications. So, the pursuit of room temperature 2D FM materials continues.   \newline
Bulk materials with soichiometry $Cr_{1-\delta}$Te, Cr$_{2}$Te$_{3}$, Cr$_{3}$Te$_{4}$ and Cr$_{5}$Te$_{6}$ are known for their magnetic properties for very long time \cite{magn5,magn6,magn7}. Although almost all  the magnetic moment is concentrated on Cr atoms, the magnetic structure of these compounds is far from being simple. Theoretical \cite{magn8} and experimental \cite{magn9} studies indicate important role of on-site Coulomb correlations as well as exchange bias which, in case of Cr$_{2}$Te$_{3}$ stems from the vacancy Cr layer. This triggered further studies on magnetism in chromium telluride nanocrystals \cite{magn10,magn11}, thin films \cite{magn12,magn13,magn14,magn15} and quasi-2D layers \cite{magn16}. 
One of the most promising candidates for pure 2D magnetic material is the family of 2D hexagonal Cr$_3$N$_{4}$ (N=S, Se, Te). According to recent theoretical predictions, 2D Cr$_{3}$Se$_{4}$ and Cr$_4$Te$_{5}$ can have T$_{C}$ as high as, respectively, 370K and 460K \cite{magn17}.
\newline
In this paper, based on $\it{ab-initio}$ calculations, we present a new 2D magnetic material with stoichiometry Cr$_{2}$Te$_{3}$, consisting of two layers of Cr atoms. We also investigate the electronic and magnetic properties of a new family of magnetic compounds M$_{2}$N$_{3}$.

\section{Calculation method}
All the calculations were conducted within Density Functional Theory (DFT) with the help of several tools. The determination of atomic and electronic properties of all structures was performed with the use of the VASP code \cite{VASP1} and the PAW-PBE (GGA) \cite{paw,pbe1} pseudopotentials with the inclusion of intra-atomic interactions on the LDA+U level \cite{LDAU}. The U values of Cr, Mn, and Ti d orbitals were set as
3.5, 3.9, and 5.0 eV, respectively. These values were chosen following those reported in the literature \cite{LDAU1,LDAU2}. 
To determine the band gap width more accurately than the GGA+U level allows, the G$_{0}$W$_{0}$ method was applied, as implemented in VASP \cite{gw}.
All the structures were optimized until the forces exerted on atoms were smaller than 10$^{-5}$ eV/$\angstr$. The dense 30 $\times$ 30 $\times$ 1 k-points uniform grid was applied.
To assess the dynamic stability of the structures, the phonon dispersion bands were determined using the frozen-phonon approach \cite{phonons1} as implemented in phonopy code \cite{phonons2}.
Since the atomic structure of Cr$_{2}$Te$_{3}$ was not known a priori, the evolutionary search was
conducted. The NGOpt code \cite{ngopt} was used for this part, which uses approach combining the 
neural networks and evolutionary techniques together with ab-initio calculations. In this case, the VASP was applied as the total energy calculator. For the estimation of the Curie temperature, the classical Monte Carlo method was used, as implemented in the Vampire code \cite{vampire1, vampire2}. 
 
\section{Results}
We started with the determination of the atomic structure of  Cr$_{2}$Te$_{3}$  compound. In NGOpt one gives as an input stoichiometry, population size, and the total number of generations. In this case, a population of 48 individuals (i.e. probe structures) per generation was chosen. The lowest energy structure was obtained as early as in the 5th generation and prevailed for 10 next generations, so it was accepted as the global minimum. The resulting structure has a hexagonal unit cell with a lattice constant equal to 3.94 $\angstr$. The Cr atoms form two layers with three layers of Te atoms, alternately (Fig. 1). 

\begin{figure}[ht]
  \begin{center}
    \begin{tabular}{c}
      \resizebox{75mm}{!}{\includegraphics[angle=0]{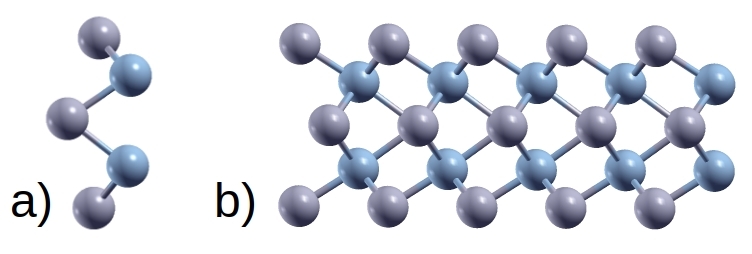}} \\
      \resizebox{75mm}{!}{\includegraphics[angle=0]{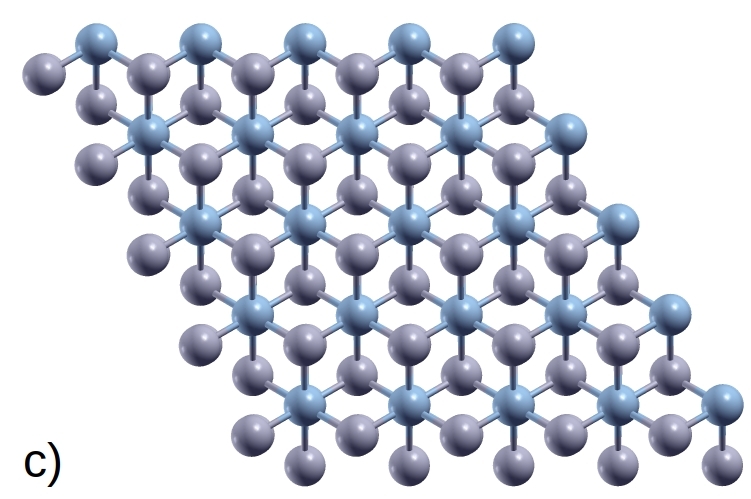}}      
    \end{tabular}
    \vspace*{-1mm}
    \caption{(Color online) Atomic structure of  Cr$_{2}$Te$_{3}$  compound: a) unit cell as seen from y-z plane, b) supercell 5$\times$5$\times$1 (x-z plane), c) supercell 5$\times$5$\times$1 (x-y plane). Cr atoms - blue, Te atoms - grey.}  
    \label{fig1}
  \end{center}
\end{figure}

Distance between Cr layers $\Delta_{Cr-Cr}$ is equal to 3.21 $\angstr$. With the total energy of -5.91 eV/atom and the phonon spectrum free of imaginary frequencies (Fig. 2b), the structure can be considered as stable. It is a semiconductor structure, with FM alignment of magnetic moments, located almost distinctly on Cr atoms. The total magnetic moment is equal to 6.00 $\mu_{B}$ per formula unit. \newline
\begin{figure*}[ht]
  \begin{center}
    \begin{tabular}{cc}
      \resizebox{90mm}{!}{\includegraphics[angle=0]{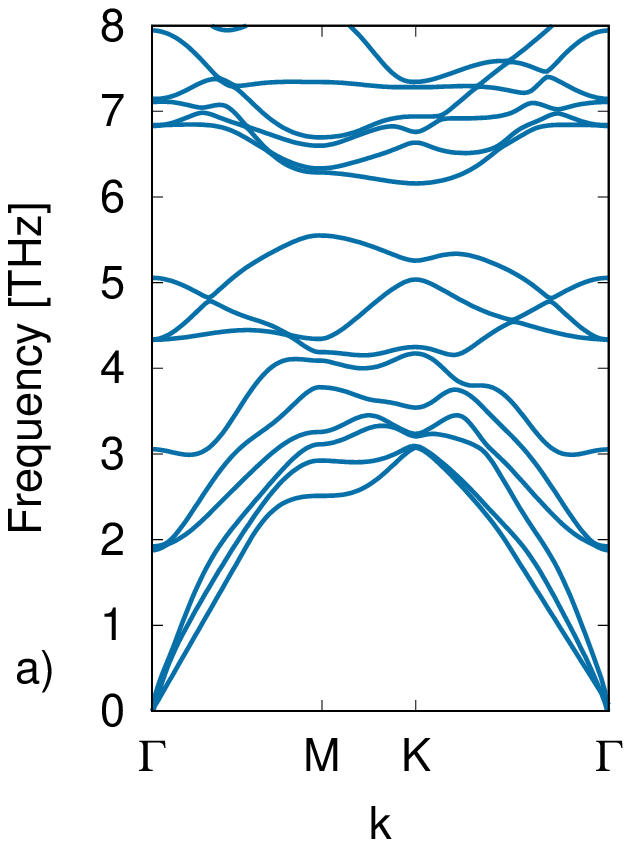}} &
      \resizebox{90mm}{!}{\includegraphics[angle=0]{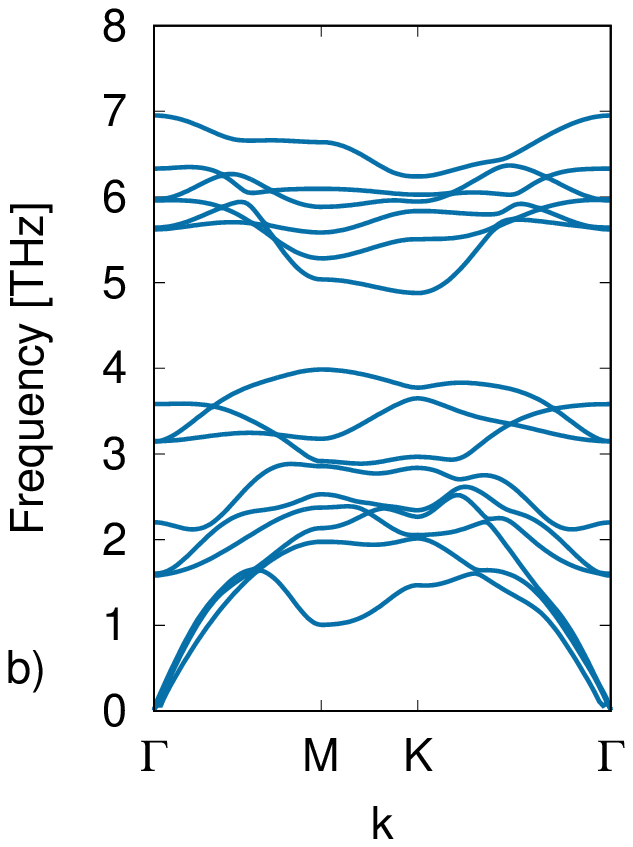}} \\    
      \resizebox{90mm}{!}{\includegraphics[angle=0]{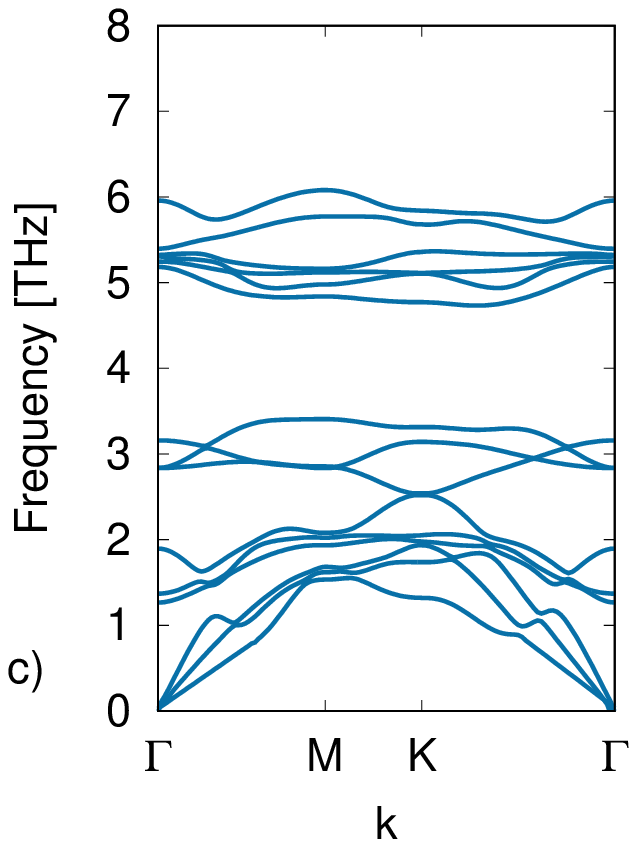}} &
      \resizebox{90mm}{!}{\includegraphics[angle=0]{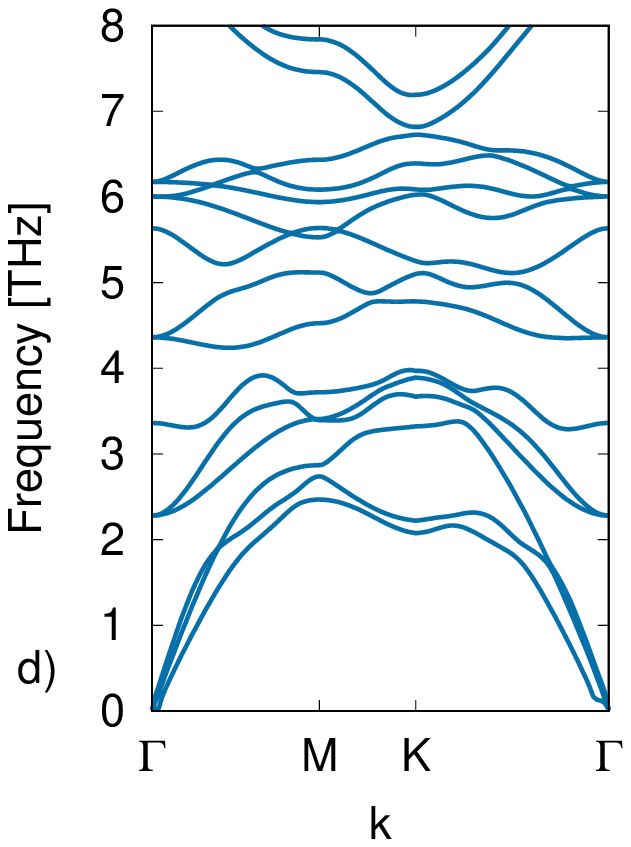}} \\   
    \end{tabular}
    \vspace*{-1mm}
    \caption{(Color online) Calculated phonon spectrum for a)  Cr$_{2}$Se$_{3}$ , b)  Cr$_{2}$Te$_{3}$ , C) Mn$_{2}$Te$_{3}$ , d) Ti$_{2}$Se$_{3}$ .}
    \label{fig2}
  \end{center}
\end{figure*}
Since the compound  Cr$_{2}$Te$_{3}$  is stable and magnetic, we checked also other stoichiometries with structure type  M$_{2}$N$_{3}$. This gives the total number of 30 structures, among which 11 turned out to be magnetic (all FM) and 4 of them stable (cf. Table I). The phonon spectra for all four stable compounds can be seen in Fig. 2. With the total energy per atom of -6.99 eV, the Ti$_{2}$Se$_{3}$  is the most bound structure among all calculated.\newline
\begin{table}[]
\begin{tabular}{|l|c|c|c|c}
\hline
Structure & a [\angstr] & E$_{tot}$/at. [eV] & Dyn. stability    \\
\hline
Ti$_{2}$Se$_{3}$  & 3.58 & -6.99& +    \\
V$_{2}$Se$_{3}$   & 3.66 & -6.36& -    \\
V$_{2}$Te$_{3}$   & 3.57 & -5.74& -    \\
Cr$_{2}$Se$_{3}$  & 3.63 & -6.45& +    \\
Cr$_{2}$Te$_{3}$  & 3.94 & -5.91& +    \\
Mn$_{2}$S$_{3}$   & 3.48 & -6.63& -    \\
Mn$_{2}$Se$_{3}$  & 3.70 & -6.16& -    \\
Mn$_{2}$Te$_{3}$  & 4.01 & -5.67& +    \\
Fe$_{2}$S$_{3}$   & 3.28 & -6.10& -    \\
Fe$_{2}$Se$_{3}$  & 3.49 & -5.57& -    \\
Fe$_{2}$Te$_{3}$  & 3.73 & -4.73& -   
\end{tabular}
\label{tab1}
\caption{Structural properties of  M$_{2}$N$_{3}$  compounds where a - lattice constant, E$_{tot}$/at. - total energy per atom and dynamic stability according to calcualted phonon spectrum.}
\end{table}
Having established the stability study, the next step was to calculate the electronic band 
structure for four stable compounds. The DFT PBE+U results can be seen in Fig. 3a–d,
showing, that the  Cr$_{2}$Se$_{3}$  and  Cr$_{2}$Te$_{3}$   structures are semiconductors with a non-direct band
gaps ($\Gamma$-K) of, respectively 0.14 and 0.16 eV, while Ti$_{2}$Se$_{3}$  and Mn$_{2}$Te$_{3}$  are metallic.
Application of the G$_{0}$W$_{0}$ method corrects the band gap widths of  Cr$_{2}$Se$_{3}$  and
 Cr$_{2}$Te$_{3}$  structures to 0.28 eV and 0.31 eV, respectively, while the other two remain
metallic (Table II). \newline
In the case for example of a graphene monolayer, the bands constituting the Dirac
cone are composed almost exclusively of the p$_{z}$ orbitals of carbon atoms.
Unlike graphene, the band structure of the hexagonal  M$_{2}$N$_{3}$  compounds
close to the Fermi level is more complex. Let us concentrate on the
semiconductor band structure of   Cr$_{2}$Te$_{3}$. Fig. 3 e,f shows the orbital character of each band. 
The band character is calculated by projecting the wave functions onto spherical harmonics
of $p$ or $d$ ($p$$_{x}$, $p$$_{y}$ $p$$_{z}$, $d$$_{xy}$, $d$$_{yz}$, $d$$_{3r^{2}-z^2}$, $d$$_{xz}$, $d$$_{x^{2}-y^{2}}$) type, which are nonzero within a sphere close to each ion \cite{VASP1}. 
As can be seen, the vicinity of the Fermi level is dominated by the $p$$_{x}$ and $p$$_{y}$ orbitals of the Te atoms (top of the valence band) and the $d$$_{x^{2}-y^{2}}$ orbitals of the Cr atoms (bottom of the conductions band). The situation is similar in the case of Cr$_{2}$Se$_{3}$. For the other two materials (Mn$_{2}$Te$_{3}$ and Ti$_{2}$Se$_{3}$) the $d$ states in the vicinity of the Fermi level (Figs. 3 c,d) are lying lower 
in energy making these structures metallic. \newline 
To analyze the magnetic properties of a 2D material, one has to be very cautious. According to the famous Mermin-Wagner theorem \cite{M-W}, magnetic moments of a 2D system of localized spins in T$\ne$0 will always
become evenly distributed (''up'' vs. ''down'') and sum up to zero total magnetic moments, if the only interaction between the sites is the exchange term. So, we started our analysis with the determination of the magnetic anisotropy energy ($MAE$) defined as E$_{MAE}$ = E$_{tot}^{\perp}$ - E$_{tot}^{\parallel}$, where E$_{tot}^{\parallel (\perp)}$ is the total energy of the system with parallel (perpendicular) alignment of magnetic moments with respect to the plane. With this definition, E$_{MAE}$ is smaller than zero for the cases where the ground state corresponds to the perpendicular alignment of magnetic moments. As can be seen from Tab. II, this is the case for three among the structures considered, except for Ti$_{2}$Se$_{3}$. The $z$ axis is then the easy axis, which means that the anisotropy may play an important role in maintaining the magnetic moment in non-zero temperatures. \newline
To estimate the Curie temperature, we used a simple model Hamiltonian: 
\beq
 H = -\frac{1}{2} J \sum_{<ij>} \vec{S_{i}} \cdotp \vec{S_{j}} 
     -A \sum_{i} (S_{i}^{2})^2 
     - \frac{1}{2} B \sum_{<ij>} S_{i}^{2} S_{j}^{2}
\eeq
where $J$,$A$, and $B$ are exchange interactions, single-ion anisotropy, and nearest neighbor anisotropic exchange coupling constants, respectively. The summation is done over the nearest neighbors for all magnetic sites. 
The projection of the DFT results onto the above Hamiltonian allowed to calculate the coupling constants (Tab. II).
\newline
To estimate the T$_{C}$, the calculated exchange ($J$) and anisotropy ($A$) couplings were put to the Vampire code, which treats the calculated magnetic moments in Eq. 1 as the classical vectors. The resulting Curie temperatures are not very high (Tab. II), but can be verified by the experimental measurements. \newline
The $J$ coupling constant in Hamiltonian (1) is responsible for the exchange interaction between Cr atoms in the unit cell. 
The distance between metallic ions (Cr, Mn, Ti) in all unit cells is too big, that this exchange can be direct. This suggests, that the interaction is mediated via the non-magnetic atom, which lies in the middle of the layer (Fig. 1b)
The Goodenough-Kanamori rules suggest the superexchange interaction between ions with an angle of 90$^{\circ}$ between them can give effective (weak) FM coupling \cite{G-K}. 
As can be seen from Fig. 1a, for  Cr$_{2}$Te$_{3}$, the angle Cr-Te-Cr is equal to 73$^{\circ}$, which is not very close to 90$^{\circ}$, so the situation is more complex. To shed some light on the nature of Cr-Cr exchange we calculated the per-orbital contributions of exchange parameter for  Cr$_{2}$Te$_{3}$  using the Green function approach, as described in \cite{korotin} 
Table III shows the calculated coupling between $d$ states of Cr atoms in the unit cell. As can be seen, the biggest FM contributions come from d$_{xz}$-d$_{xz}$ and d$_{yz}$-d$_{yz}$ couplings. Moreover, there are also AFM couplings, originating mainly from d$_{3r^{2} - z^2}$ orbitals. Off-diagonal terms are of less importance. These results suggest, that the superexchange interaction in  Cr$_{2}$Te$_{3}$  can be attributed to FM coupling mediated by the $\sigma$ Te orbitals with the small admixture of AFM coupling mediated by the $\pi$ Te orbitals. For the other three compounds, the FM/AFM couplings ratio is a little different, but generally, this mechanism is also valid.  
\begin{table*}[]
\begin{tabular}{|l|c|c|c|c|c|c|c|c|c} 
\hline
Structure & U [eV] & E$_{gap}$ PBE+U (G$_{0}$W$_{0}$) [eV] & Magn. mom. [$\mu_{B}$] & MAE [meV] &J [meV]  & A [meV] & B [meV] & T$_{C}$ [K]  \\
\hline
Ti$_{2}$Se$_{3}$  & 5.0 & m           & 1.62 &  1.23 & - & - & - & -   \\
Cr$_{2}$Se$_{3}$  & 3.5 & 0.14 (0.28) & 6.00 & -0.23 & 0.163 & 0.029 & -0.003 & 30   \\
Cr$_{2}$Te$_{3}$  & 3.5 & 0.16 (0.31) & 6.00 & -0.55 & 2.515 & 0.134 & -0.047 & 55  \\
Mn$_{2}$Te$_{3}$  & 3.9 & m           & 7.48 & -2.83 & 1.843 & 0.098 & -0.034 & 45  
\end{tabular} 
\label{tab2}
\caption{Electronic and magnetic properties of  M$_{2}$N$_{3}$  compounds: U - constant for PBE+U approx., band gap calculated on the PBE (G$_{0}$W$_{0}$) level of approx., the magnetic moment per unit cell, $MAE$ - magnetic anisotropy energy, $J$, $A$, $B$ -  constants of the Hamiltonian (1), Curie temperature estimated by classical MD simulation.}
\end{table*}
\begin{table}[]
\begin{tabular}{|l|r|r|r|r|r|}
\hline
   $J$$_{i,j}$          & $3z^2-r^2$ & $xz$ & $yz$ & $x^2-y^2$ & $xy$ \\ \hline
$3z^2-r^2$   &  0.70951 & -0.02434 &  0.06560 &  0.00858 & -0.00681      \\ \hline
$xz$         & -0.02434 & -1.23703 & -0.02792 &  0.05350 &  0.17230      \\ \hline
$yz$         &  0.06560 & -0.02792 & -1.18891 &  0.17664 &  0.05476      \\ \hline
$x^2-y^2$    &  0.00858 &  0.05350 &  0.17664 &  0.05170 & -0.04649      \\ \hline
$xy$         & -0.00681 &  0.17230 &  0.05476 & -0.04649 &  0.03423      \\ \hline
\end{tabular} 
\label{tab3}
\caption{Calculated per-orbital exchange parameters for Cr$_{2}$Te$_{3}$ , $i$,$j$ - $d$ orbitals of the Cr atoms.}
\end{table}
\begin{figure*}[h!] 
  \begin{center}
    \begin{tabular}{cc} 
      \resizebox{90mm}{!}{\includegraphics[angle=0]{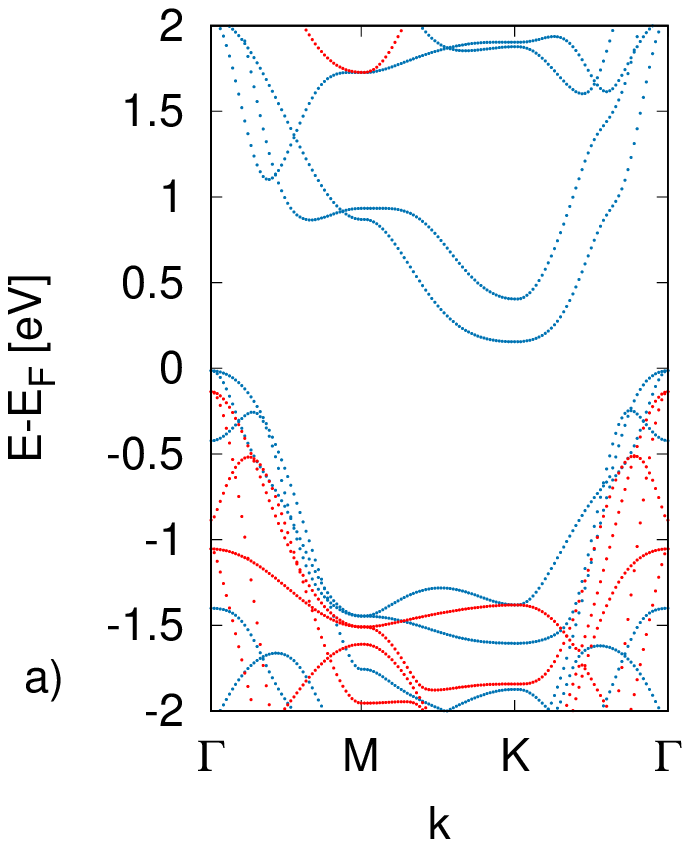}} &
      \resizebox{90mm}{!}{\includegraphics[angle=0]{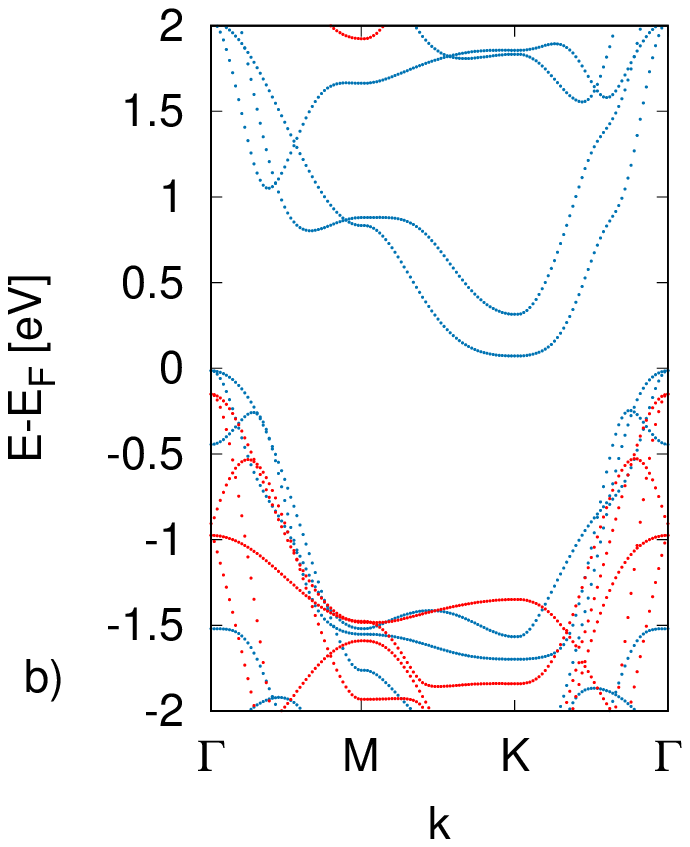}} \\     
      \resizebox{90mm}{!}{\includegraphics[angle=0]{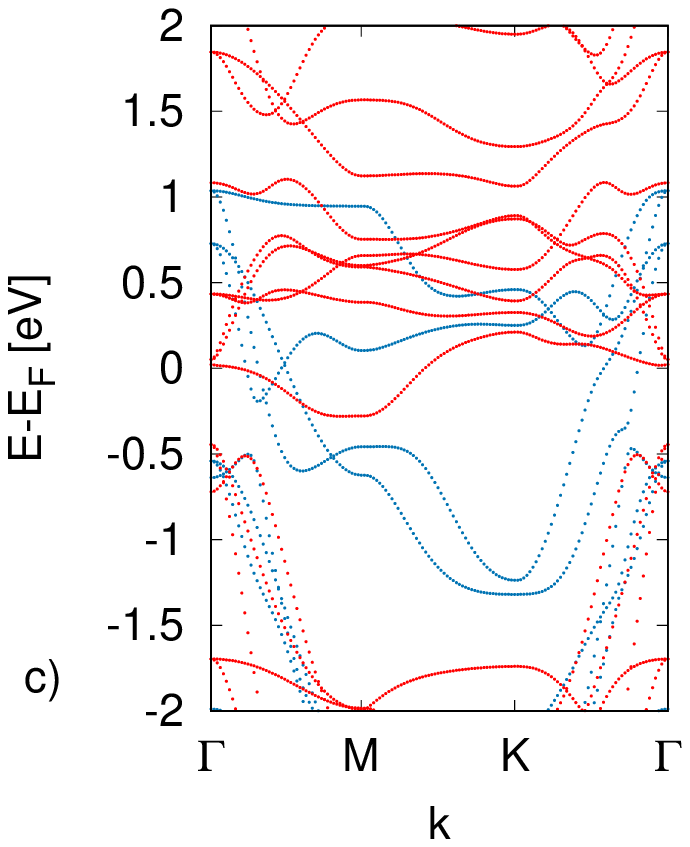}} &
      \resizebox{90mm}{!}{\includegraphics[angle=0]{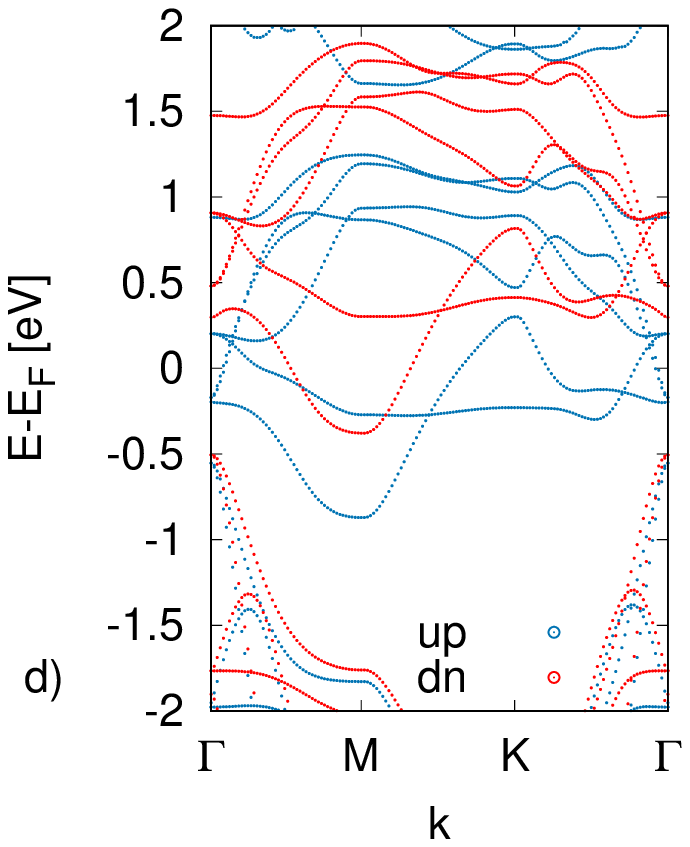}} \\ 
      \resizebox{90mm}{!}{\includegraphics[angle=0]{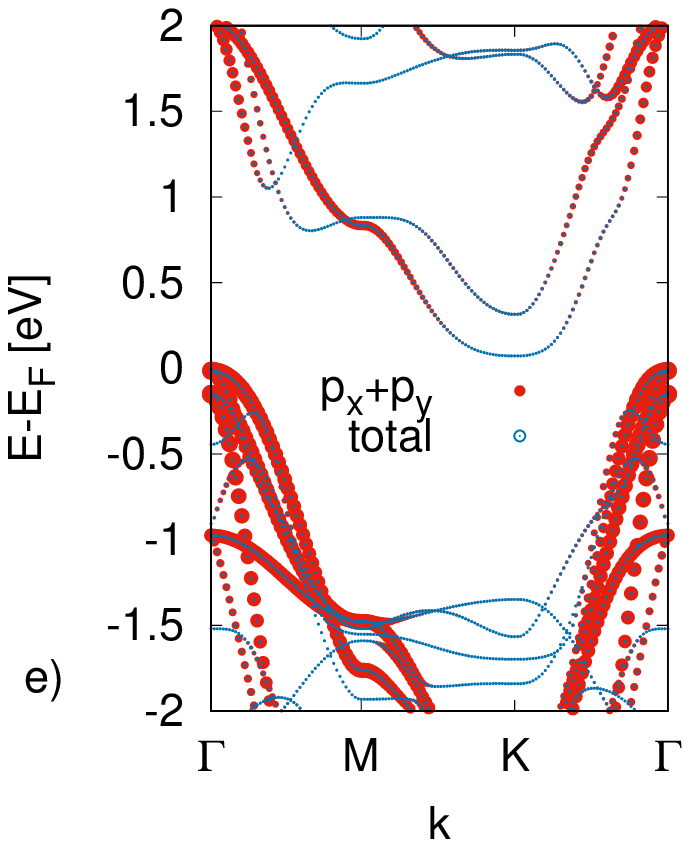}} &
      \resizebox{90mm}{!}{\includegraphics[angle=0]{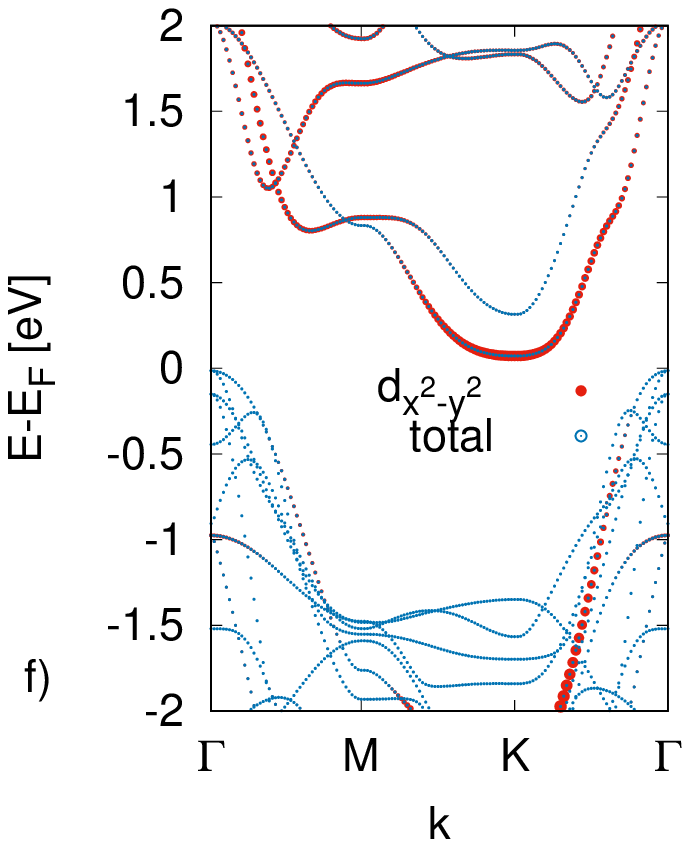}}   
    \end{tabular}
    \vspace*{-1mm}
    \caption{(Color online) Calculated spin-resolved electronic band structure spectrum for a)  Cr$_{2}$Se$_{3}$ , b)  Cr$_{2}$Te$_{3}$ , C) Mn$_{2}$Te$_{3}$ , d) Ti$_{2}$Se$_{3}$ , e,f) orbital character of bands for  Cr$_{2}$Te$_{3}$ .}  
    \label{fig3}
  \end{center}
\end{figure*}

\section{Conculsions}
\textit{Ab-initio} calculations performed for Cr$_2$Se$_3$, Cr$_2$Te$_3$ and Mn$_2$Te$_3$ 2D monolayer structures clearly show that such systems are stable and exhibit interesting magnetic properties.
The magnetic moments are high (6.00 to 7.48 $\mu_{B}$ per formula unit) and the $MAE$ values suggest that their alignment is stably perpendicular to the plane. Moreover, the exchange mechanism is non-trivial, 
combining FM and AFM coupling of $d$ orbitals of metal atoms mediated by the $p$ orbitals of the non-metal,
dependent on their symmetry. \newline
The presented results suggest, that the monolayer M$_2$N$_3$ structures can be considered as potential candidates for applications in spintronic devices.

\begin{acknowledgments}
Numerical calculations were supported in part by PL-Grid Infrastructure. 
\end{acknowledgments}


\begin{thebibliography}{0}
\expandafter\ifx\csname natexlab\endcsname\relax\def\natexlab#1{#1}\fi
\expandafter\ifx\csname bibnamefont\endcsname\relax
  \def\bibnamefont#1{#1}\fi
\expandafter\ifx\csname bibfnamefont\endcsname\relax
  \def\bibfnamefont#1{#1}\fi
\expandafter\ifx\csname citenamefont\endcsname\relax
  \def\citenamefont#1{#1}\fi
\expandafter\ifx\csname url\endcsname\relax
  \def\url#1{\texttt{#1}}\fi
\expandafter\ifx\csname urlprefix\endcsname\relax\def\urlprefix{URL }\fi
\providecommand{\bibinfo}[2]{#2}
\providecommand{\eprint}[2][]{\url{#2}}

\end{thebibliography}


\begin{thebibliography}{99}
%\bibitem{add} .

\bibitem{materi2D} Briggs, S. Subramanian, Z. Lin, X. Li, X. Zhang et. al., 2D Mater. 6, 022001 (2019).
\bibitem{magn1} C. Gong, L. Li, Z. Li, H. Ji, A. Stern, Y. Xia, T. Cao, W. Bao, C. Wang, Y. Wang, Z. Q. Qiu, R. J. Cava, S. G. Louie, J. Xia and X. Zhang, Nature, 546, 265 (2017).
\bibitem{magn2} B. Huang, G. Clark, E. Navarro-Moratalla, D. R. Klein, R. Cheng, K. L. Seyler, D. Zhong, E. Schmidgall, M. A. McGuire, D. H. Cobden, W. Yao, D. Xiao, P. Jarillo-Herrero and X. Xu, Nature
546, 270 (2017).
\bibitem{magn3}. Z. Fei, B. Huang, P. Malinowski, W. Wang, T. Song, J. Sanchez, W. Yao, D. Xiao, X. Zhu, A. F. May, W. Wu, D. H. Cobden, J. H. Chu and X. Xu, Nat. Mater., 17, 778 (2018).
\bibitem{magn5} A. F. Andersen, Acta Chem. Scand. 24, 3495 (1970).
\bibitem{magn6} J. Dijkstrat, H. H. Weitering'i, C. F. van Bruggen, C. Haast and R. A.
de Groot, J. Phys.: Condens. Matter 1 9141 (1989).
\bibitem{magn7} K. Sato, Y. Aman, M. Hirai, and M. Fujisawa, J. Phys. Soc. Jpn. 59, 435
(1990).
\bibitem{magn8} S. J. Youn, S. K. Kwon and B. I. Min, J. App. Phys. 101, 09G522 (2007).
\bibitem{magn9} Lu Hui, S. T. Lim, J. F. Bi, and K. L. Teo,J. App. Phys. 111, 07D719 (2012).
\bibitem{magn10} K. Ramasamy, D. Mazumdar, R. D. Bennett and A. Gupta, Chem. Commun., 48, 5656 (2012).
\bibitem{magn11} F. Wang,J. Du, F. Sun, R. F. Sabirianov, N. Al-Aqtash, D. Sengupta, H. Zeng and X. Xu,
Nanoscale 10, 11028 (2018).
\bibitem{magn12} H. Li, L. Wang, J. Chen, T. Yu, L. Zhou, Y. Qiu, H. He, F. Ye, I. Keong Sou and G. Wang,
ACS Appl. Nano Mater. 2, 6809 (2019).
\bibitem{magn13} M. Burn, L. B. Duffy, R. Fujita, S. L. Zhang, A. I. Figueroa, J. Herrero-Martin,
G. van der Laan and T. Hesjedal, Scientific Reports 9, 10793 (2019).
\bibitem{magn14} A. Roy, S. Guchhait, R. Dey, T. Pramanik, Ch.-Ch. Hsieh, A. Rai and
S. K. Banerjee, ACS Nano 9, 4, 3772 (2015).
\bibitem{magn15} R. Akiyama, H. Oikawa, K. Y., and S. Kuroda, Phys. Status Solidi C 11, 7-8, 1320 (2014).
\bibitem{magn16} M. Bester, I. Stefaniuk and M. Kuzma, Acta Phys. Pol. A, 127, 433 (2015).
\bibitem{magn17} X. Zhang, B. Wang, Y. Guo, Y. Zhang, Y. Chen and J. Wang, Nanoscale Horiz., 4, 859 (2019)
%
\bibitem{VASP1} G. Kresse and J. Hafner, Phys. Rev. B, 47:558, (1993); G. Kresse and J. Hafner, Phys. Rev. B, 49:14251, (1994); G. Kresse and J. Furthm{\"u}ller, Comput. Mat. Sci., 6:15, (1996); G. Kresse and J. Furthm{\"u}ller, Phys. Rev. B, 54:11169, (1996).
\bibitem{LDAU} A Rohrbach, J Hafner and G Kresse J. Phys.: Condens. Matter 15, 979–996 (2003).
\bibitem{LDAU1} L. Wang, T. Maxisch, and G. Ceder, Phys. Rev. B 73, 195107 (2006).
\bibitem{LDAU2} A. Jain, G. Hautier, S. P. Ong, C. J. Moore, C. C. Fischer,
K. A. Persson, and G. Ceder, Phys. Rev. B 84, 045115 (2011).
\bibitem{phonons1} K. Parlinski, Z. Q. Li, and Y. Kawazoe, Phys. Rev. Lett. 78, 4063 (1997).
\bibitem{phonons2} A. Togo and I. Tanaka, Scr. Mater., 108, 1-5 (2015).
\bibitem{gw}  F. Fuchs, J. Furthm{\"u}ller, F. Bechstedt, M. Shishkin, and G. Kresse, Phys. Rev. B 76, 115109 (2007).
\bibitem{ngopt} K. Zberecki, arXiv:2005.06847 (2020).
\bibitem{paw} P. E. Blochl, Phys. Rev. B, 50:17953, (1994); G. Kresse and D. Joubert, Phys. Rev. B, 59:1758, (1999).
\bibitem{pbe1} J. P. Perdew, K. Burke, and M. Ernzerhof, Phys. Rev. Lett. 77, 3865 (1996), doi:10.1103/PhysRevLett.77.3865.
\bibitem{vampire1} R. F. L. Evans, W. J. Fan, P. Chureemart, T. A. Ostler, M. O. A. Ellis and R. W. Chantrell
J. Phys.: Condens. Matter 26, 103202 (2014).
\bibitem{vampire2} P. Asselin, R. F. L. Evans, J. Barker, R. W. Chantrell, R. Yanes, O. Chubykalo-Fesenko, D. Hinzke, and U. Nowak Phys. Rev. B 82, 054415 (2010). 
\bibitem{vdw} S. Grimme, J. Comp. Chem. 27, 1787 (2006).
\bibitem{M-W} N. D. Mermin, H. Wagner, Phys. Rev. Lett. 17, 22 (1966).
\bibitem{G-K} H. Weihe, and H. U. Gudel, Inorg. Chem.,  36 (17), 3632 (1997).
\bibitem{korotin} Dm. M. Korotin, V. V. Mazurenko, V. I. Anisimov, and S. V. Streltsov, Phys. Rev. B, 91, 224405 (2015).
\end{thebibliography}
\end{document}